\documentclass[conference]{IEEEtran}

\usepackage[letterpaper, left=1in, right=1.02in, bottom=1in, top=0.75in]{geometry}

\usepackage{times}
\usepackage{amsmath}
\usepackage{float}
\usepackage{array}
\usepackage{amssymb}
\usepackage{amsfonts}
\usepackage{graphicx}
\usepackage{epstopdf}
\usepackage{slashed,graphicx}
\usepackage{url}
\usepackage{tablefootnote}
\usepackage{color}
\usepackage{xcolor}
\usepackage{cite}

\ifCLASSOPTIONcompsoc \usepackage[caption=false,font=normalsize,labelfon
t=sf,textfont=sf]{subfig}
\else
\usepackage[caption=false,font=footnotesize]{subfig}
\fi

\ifodd 1
\newcommand{\rev}[1]{{{\color{black} #1}}} 
\newcommand{\revv}[1]{{{\color{black} #1}}} 
\fi

\IEEEoverridecommandlockouts

\begin{document}
%
\title{Detecting LTE-U Duty Cycling Misbehavior for Fair Sharing with Wi-Fi in Shared Bands}

\author{\IEEEauthorblockN{Xuhang Ying}
\IEEEauthorblockA{Electrical Engineering \\ University of Washington \\ Seattle, WA 98195 \\
Email: xhying@uw.edu}
\and
\IEEEauthorblockN{Radha Poovendran}
\IEEEauthorblockA{Electrical Engineering \\ University of Washington \\Seattle, WA 98195 \\
Email: rp3@uw.edu}
\and
\IEEEauthorblockN{Sumit Roy}
\IEEEauthorblockA{Electrical Engineering \\ University of Washington \\Seattle, WA 98195 \\
Email: sroy@uw.edu}

\thanks{ This work was supported in part by NSF CNS 1617153 and NSF CPS award CNS 1446866.}

}


%


\maketitle

\begin{abstract}
Coexistence of Wi-Fi and LTE Unlicensed (LTE-U) in shared or unlicensed bands has drawn growing attention from both academia and industry.
An important consideration is fairness between Wi-Fi and duty cycled LTE-U, which is often defined in terms of channel access time, as adopted by the LTE-U Forum. 
Despite many studies on duty cycle adaptation design for fair sharing, one crucial fact has often been neglected: LTE-U systems unilaterally control LTE-U duty cycles; hence, as self-interested users, they have incentives to misbehave, e.g., transmitting with a larger duty cycle that exceeds a given limit, so as to gain a greater share in channel access time and throughput. 
In this paper, we propose a scheme that allows the spectrum manager managing the shared bands to estimate the duty cycle of a target LTE-U cell based on PHY layer observations from a nearby Wi-Fi AP, without interrupting normal Wi-Fi operations. 
We further propose a thresholding scheme to detect duty cycling misbehavior (i.e., determining if the duty cycle exceeds the assigned limit), and analyze its performance in terms of detection and false alarm probabilities. 
The proposed schemes are implemented in \textit{ns3} and evaluated with extensive simulations. 
Our results show that the proposed scheme provides an estimate within $\pm$ 1\% of the true duty cycle, and detects misbehavior with a duty cycle 2.8\% higher than the limit with a detection probability of at least 95\%, while keeping the false alarm probability less than or equal to 1\%. 
\end{abstract}


\IEEEpeerreviewmaketitle

\section{Introduction}
The exponential growth of data services, such as mobile multimedia and Internet applications on portable devices like smartphones and tablets, has translated into a proportionate surge in demand for additional wireless network capacity. 
One solution that has received great attention from both academia and industry is to deploy existing technologies (e.g., small cells) in \textit{shared or unlicensed bands} to enable more efficient spectrum utilization and provide greater broadband capacity. 
To date, the FCC has opened up several bands for broadband access, 
including TV white spaces (TVWS) 
\cite{fcc2008second}, the 2.4GHz and 5GHz unlicensed bands for  proposed unlicensed LTE operations \cite{3gpp2014summary}, and the Citizens Broadband Radio Service (CBRS) bands in 3.5GHz \cite{fcc2012}.


In the current paradigm of shared spectrum, there typically exists a spectrum manager (e.g., geo-location database in TVWS or spectrum access server in  CRBS bands) that manages shared bands.
Unlicensed devices and networks operating in shared bands usually have equal spectrum access priorities, and are expected to \textit{coexist} in the same frequency, time and space. 
It is \rev{widely believed} that Wi-Fi and LTE are among the most dominant technologies that will be deployed in shared bands in the next few years, which have different channel access mechanisms.
With CSMA/CA\footnote{Carrier-Sense Multiple Access with Collision Avoidance}, each Wi-Fi device senses the medium, and allows others to finish transmission before attempting its own transmission, while LTE transmits continuously without sensing, as it traditionally operates exclusively in bands owned by operators.
As a result, LTE would block Wi-Fi transmissions during coexistence, resulting in degraded Wi-Fi performance \cite{cavalcante2013performance}.

In order to achieve time-division-multiplexing (TDM) based coexistence with Wi-Fi, two types of LTE have been proposed for unlicensed operations: LAA and LTE-U. 
The former employs listen-before-talk (LBT) mechanisms, 
\rev{while the latter} is duty-cycle based and proposed for supplementary downlink. 
LTE-U exploits existing LTE functionality (e.g., almost blank subframes \cite{almeida2013enabling}) to create alternating ON/OFF periods so as to accommodate Wi-Fi transmissions. 
In this paper, we are interested in LTE-U that is intended for earlier commercialization in markets where regulations do not require LBT, such as China, Korea, India and the USA. 

Wi-Fi/LTE-U coexistence has drawn growing attention from different aspects \cite{sadek2015extending, almeida2013enabling, sagari2015coordinated}, and one important consideration is \textit{fairness}.
Since both are TDM based, one natural criterion
is fair sharing in channel access time, i.e., fraction of LTE-U ON duration in each cycle (aka. duty cycle, a quantity between $0$ and $1$) \rev{should not be} more than a limit.
For example, when coexisting with a Wi-Fi network, the LTE-U AP\footnote{Throughput the paper, LTE-U eNB (Evolved Node B) is referred to as access point (AP) for the purpose of convenience.} should not transmit for more than $50\%$ of the time. 
In fact, this criterion has already been adopted by the LTE-U Forum as part of the coexistence specifications \cite{lteUforum2015coexistence}. 

Although many researchers have studied LTE-U duty cycle adaptation design for fair sharing (e.g. \cite{qualcomm2015,rupasinghe2015reinforcement,parvez2016cbrs}), a crucial fact has often been neglected: 
\rev{Wi-Fi nodes, as benign users, can only access the channel  during LTE-U OFF time, while ON/OFF time is under unilateral control of LTE-U APs.}  
Therefore, LTE-U APs, as self-interested users, will have incentives to \textit{misbehave}, that is, \rev{transmitting with a larger duty cycle that exceeds the limit}, so as to gain a greater share in channel access time and throughput. 
This is a realistic concern (from Wi-Fi operators and users), 
especially when LTE-U operators are not likely to disclose details of their proprietary duty cycle adaptation algorithms.
Such concern persists, unless a proper fairness monitoring scheme that consists of duty cycle estimation and misbehavior detection is in place, and this need has also been acknowledged in \cite{bhattarai2016overview}.

In this work, we propose monitoring of Wi-Fi/LTE-U channel access time fairness as part of the spectrum manager's functionality and responsibility. 
Specifically, the spectrum manager assigns a reasonable duty cycle limit to a LTE-U AP, and estimates its duty cycle to see if it exceeds the assigned limit.
To the best of our knowledge, this is the first paper that discusses fairness monitoring for Wi-Fi/LTE-U coexistence in shared bands.

For the above purpose, the spectrum manager may deploy a LTE-compatible device close to the target LTE-U AP to measure LTE-U ON time and estimate its duty cycle.
But deploying such device for each LTE-U AP can be costly. 
Alternatively, an energy detector may be deployed to measure the total ON duration, and subtract the portion due to Wi-Fi activities (e.g., by detecting Wi-Fi preambles).
In fact, such energy detector is already available at each Wi-Fi device, which is able to measure ON duration with $1\mu$s granularity.
It means that the spectrum manager can collect PHY layer observations from a Wi-Fi AP close to the target LTE-U AP, and estimate its duty cycle to detect possible misbehavior.

Our primary contributions are as follows:
\begin{itemize}
\item We consider coexistence between a LTE-U cell and a Wi-Fi network, and propose a scheme that allows the spectrum manager to estimate the LTE-U duty cycle in a cycle period based on observed busy periods from a local Wi-Fi AP without interrupting normal operations of the Wi-Fi network.

\item We propose a thresolding scheme for misbehavior detection.
We analyze its detection performance in terms of detection probability $P_d$ and false alarm probability $P_{fa}$. 
Our analysis shows that smaller Wi-Fi packets and a larger LTE-U cycle period would improve $P_d$ and reduce $P_{fa}$.

\item We implement the proposed schemes in \textit{ns3} and evaluate their performance with extensive experiments. 
Our results show that for a typical LTE-U cycle period of $160$ms with a $2$ms idle gap every $20$ms ON duration \cite{qualcomm2015}, the estimation error is within $\pm 1\%$ of the true duty cycle.
Besides, the proposed scheme detects misbehavior with a duty cycle that is $2.8\%$ larger than the limit with $P_d$ at least $95\%$ and $P_{fa}$ less than or equal to $1\%$.  

\end{itemize}

The rest of this paper is organized as follows.
In Section~\ref{sec:related_work}, a brief review of related work is provided. 
In Sections~\ref{sec:system_model} and \ref{sec:monitoring}, we describe our system model, and present fairness monitoring schemes, respectively.
We evaluate the proposed schemes in Section~\ref{sec:evaluation}, and conclude this study in Section~\ref{sec:conclusion}.

\section{Related Work}\label{sec:related_work}
Wi-Fi and LTE-U coexistence is being actively studied in the recent years. 
In \cite{almeida2013enabling}, Almeida {\em et al.} implemented LTE-U duty cycles by modifying the almost blank subframe functionality. 
Such duty cycle can be static as in \cite{nihtila2013system}, 
or adaptive as in Carrier Sensing Adaptive Transmission (CSAT), which was proposed by Qualcomm \cite{qualcomm2015}.
It allows a LTE-U AP to sense and measure medium utilization during OFF time, and adjust duty cycles accordingly. 
In \cite{cano2015coexistence}, Cano \textit{et al.} proposed a duty-cycle mechanism to achieve proportional fairness among LTE-U and WiFi, by selecting an appropriate probability to
access the channel and transmission duration.
Other techniques include Q-learning \cite{rupasinghe2015reinforcement} and the multi-armed bandit approach \cite{parvez2016cbrs}, \rev{which dynamically adjust duty cycles based on channel usage}.
 
In reality, duty cycle adaptation schemes are most likely to be proprietary, and their details would not be revealed by LTE-U operators.
Thus, without a proper duty cycle estimation and misbehavior detection scheme, fair sharing between Wi-Fi and LTE-U can only be at the mercy of LTE-U operators.
In this paper, we propose a scheme that allows the spectrum manager to estimate LTE-U duty cycle, and detects possible misbehavior. 

%



\section{System Model}\label{sec:system_model}
In this section, we provide brief background on Wi-Fi, and describe the duty cycled LTE-U model.
Then we formally define channel access time fairness, and present our fairness monitoring architecture.

\subsection{Wi-Fi Basics}\label{sec:wifi_basics}
The Wi-Fi standard 
\cite{wifi2012} 
employs CSMA/CA that implements the Distributed Coordination Function (DCF) -- a distributed slotted medium access scheme with an exponential back-off. 
In DCF, each node attempting to transmit must ensure the medium has been idle for a DIFS (DCF Interframe Spacing) period \revv{(i.e., $34\mu$s)}.
Then it selects a back-off (BO) counter uniformly at random from $[0, CW-1]$, where $CW$ is the contention window with an initial value of $CW_{\min}$.
Each failed transmission doubles $CW$, up to $CW_{\max}$, and each successful transmission resets $CW$ to $CW_{\min}$.
After a DIFS idle period, the counter is reduced by one every BO slot (i.e., $9\mu$s), if no other transmissions are detected during the countdown. Otherwise, the counter is frozen until the medium is once again idle for a DIFS period. 
 
\begin{figure}[ht!]
	\begin{center}
    \includegraphics[width = 1\columnwidth]{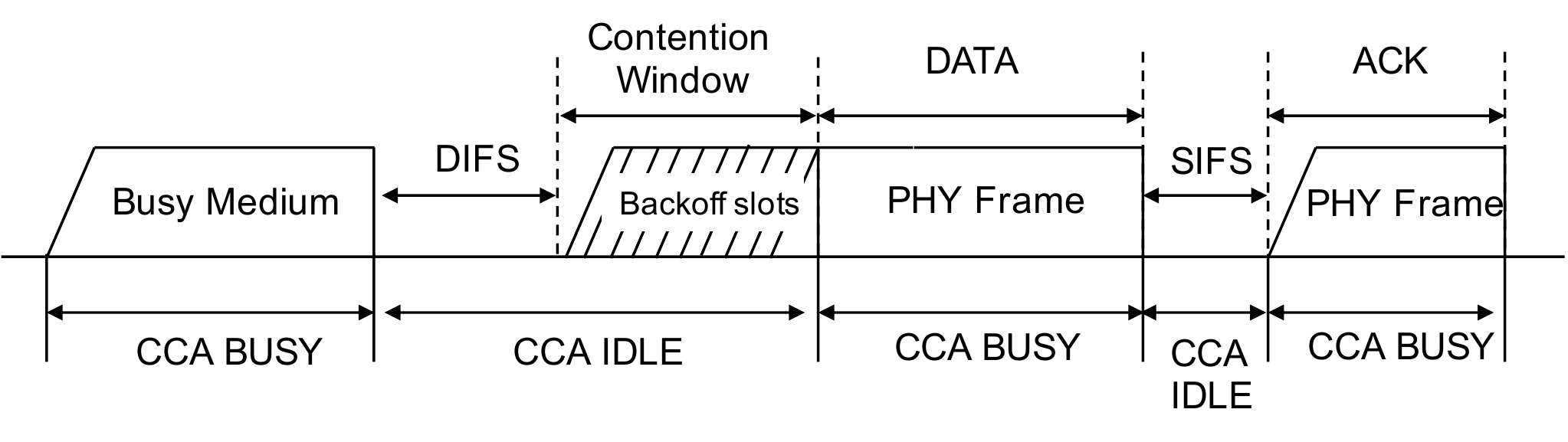}
    \caption{Typical CSMA/CA access cycle.}
    \vspace{-0.5cm}
    \label{fig:CSMA_CA_access_cycle}
    	\end{center}
\end{figure}

Each Wi-Fi node performs Clear Channel Assessment (CCA) to determine if medium is idle or busy.
It has two functions:
\begin{itemize}
\item \textit{Carrier sense} (CS): 
The ability to detect the preamble of a valid Wi-Fi transmission at a signal level equal to or greater than $-82$dBm/$20$MHz.

\item \textit{Energy detection} (ED): 
The ability to detect the energy of non-Wi-Fi transmissions
(or Wi-Fi transmissions with missed preamble) 
at a signal level equal to or greater than $-62$dBm/$20$MHz. 
\end{itemize}
A typical CSMA/CA access cycle 
is shown in Fig.~\ref{fig:CSMA_CA_access_cycle}.
When a packet is successfully received, the intended receiver will transmit an acknowledgement (ACK) after a SIFS (Short Interframe Space) period \revv{(i.e., $16\mu$s)}. 





\subsection{Duty Cycled LTE-U}\label{sec:LTE_U_model} 
\begin{figure}[ht!]
	\begin{center}
    \includegraphics[width = .8\columnwidth]{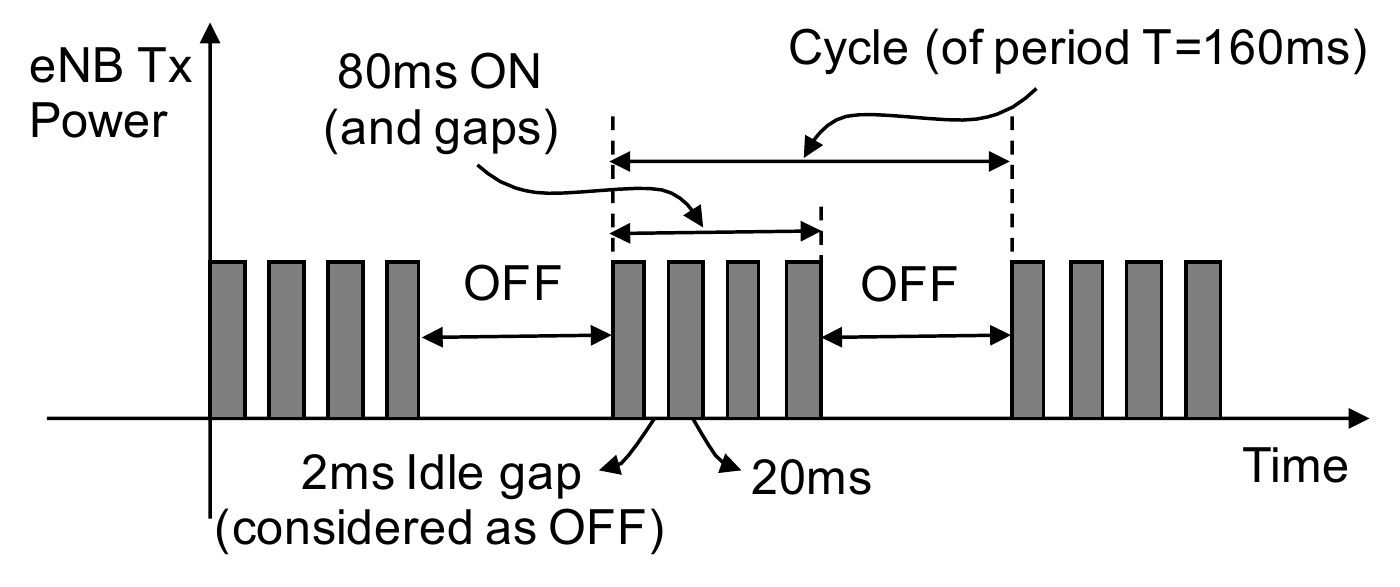}
    \caption{Example of duty cycled LTE-U with a cycle of $160$ms and a duty cycle of $0.5$ (i.e., $80/80$ms ON/OFF). 
    Idle gaps are introduced every $20$ms ON duration. 
    }
    \vspace{-0.3cm}
    \label{fig:LTE_U_duty_cycle}
    	\end{center}
\end{figure}

Fig.~\ref{fig:LTE_U_duty_cycle} illustrates the duty cycling behavior of a LTE-U AP as in CSAT \cite{qualcomm2015}.
It operates in the shared channel with a period of $T$ (ms), ranging from $10$s to $100$s of ms (typical values are $80$ms and $160$ms, or as large as $640$ms \cite{qualcomm2015}). 
In each cycle, it transmits (i.e., ON) for a fraction of time $\alpha \in (0,1)$, i.e., \textit{duty cycle}, and stays OFF in the remaining time. 
To protect Wi-Fi flush delay-sensitive data, frequent idle gaps (of few msec) are introduced in ON duration \cite{qualcomm2015} such that the maximum \textit{continuous} ON duration is no greater than a limit (e.g., $20$ms) \cite{lteUforum2015coexistence}.
We assume that the minimum continuous ON time is larger than the maximum transmission time of Wi-Fi packets.

\subsection{Fairness in Channel Access Time}
In this work, we are interested in fair sharing of channel access time between one LTE-U cell\footnote{When multiple LTE-U cells are present, they may coordinate and transmit in different portions of the same cycle. 
For instance, with two LTE-U cells $A$ and $B$, $A$ only transmits in the first half of the cycle, while $B$ only transmits in the second half. 
If so, LTE-U cells can be monitored separately. } 
and one Wi-Fi network that share one $20$MHz channel.
We assume that Wi-Fi nodes are interfered by the LTE-U AP (i.e., received LTE-U interference at each Wi-Fi node exceeds the CCA-ED threshold), and they are within each other's CS range.
It is important to note that multiple  Wi-Fi networks that are overlapping or in close range can be considered as a single larger network if we assume that no hidden nodes exist\footnote{The impact of hidden Wi-Fi nodes is deterred for future work.}.
In this case, Wi-Fi nodes will sense LTE-U interference and deter their transmissions \rev{until LTE-U ON time is over};
during OFF time, they will contend for the next transmission opportunity.

Given a cycle of period $T$, define $ON_i$ as the $i$-th \textit{continuous} ON duration, where $1\leq i\leq n$.
According to the coexistence specifications proposed by the LTE-U Forum\cite{lteUforum2015coexistence}, it is required that the duty cycle be less than or equal to a limit $\alpha_{\max}$, i.e.,
 \begin{equation}
\alpha = \frac{1}{T} \sum_{i=1}^n ON_i \leq \alpha_{\max}  \in (0,1). \label{eq:duty_cycle}
\end{equation}
For instance, if a LTE-U cell is coexisting with a Wi-Fi network, $\alpha_{\max}$ may be set to $50\%$;
with two Wi-Fi networks, $\alpha_{\max}$ may be set to $33\%$ instead.

\begin{figure}[ht!]
	\begin{center}
    \includegraphics[width = .7\columnwidth]{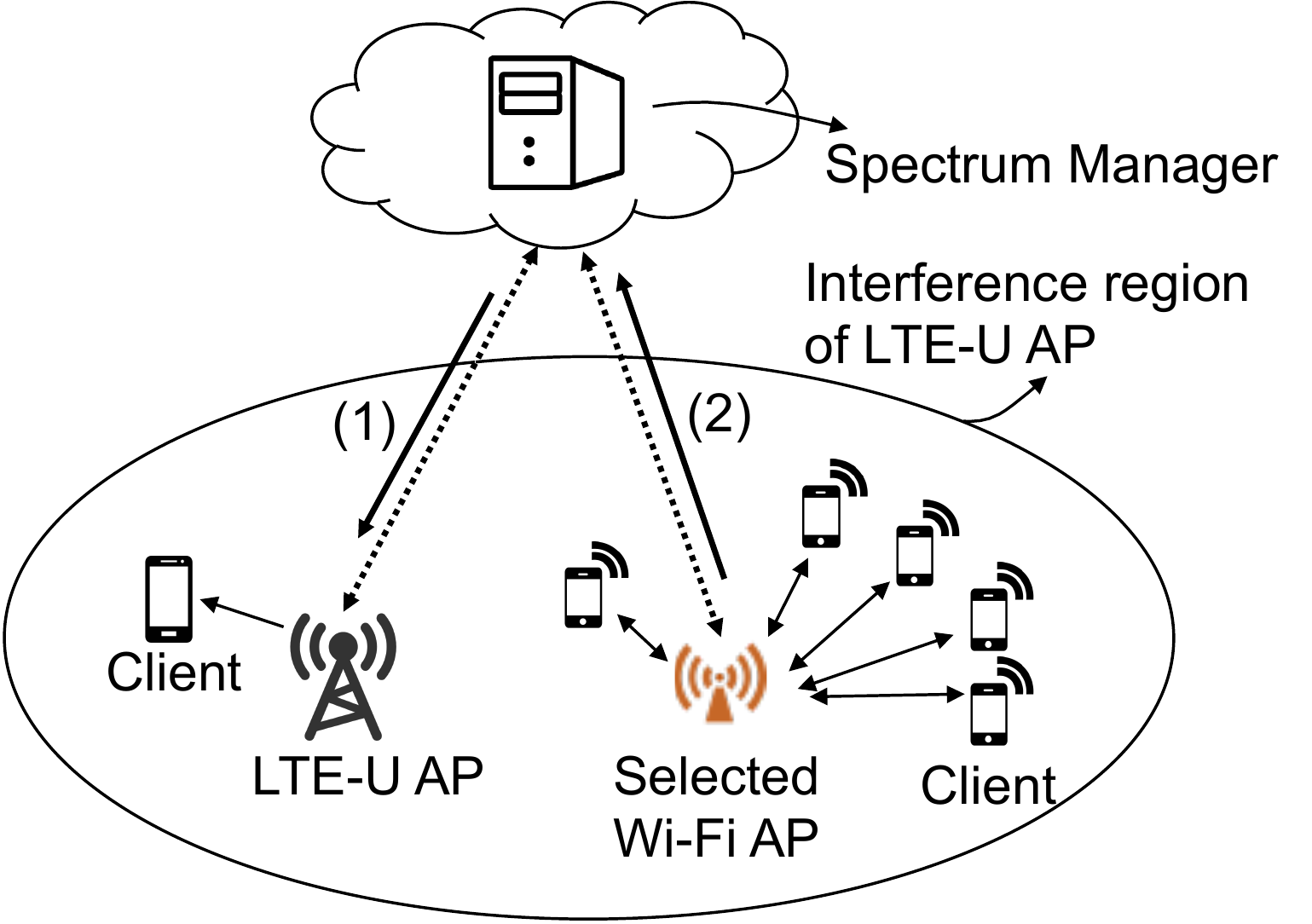}
    \caption{Co-channel deployment of a LTE-U cell and a Wi-Fi network. 
	(1) The spectrum manager determines a reasonable duty cycle limit for the LTE-U AP. 
	(2) Then it gathers PHY layer observations from a nearby Wi-Fi AP, so as to estimate the duty cycle of the LTE-U AP and detect any misbehavior.
    }
    \label{fig:scenario}
    	\end{center}
\end{figure}

\subsection{Fairness Monitoring}
In shared bands (e.g., TVWS and CBRS bands), there exists a spectrum manager that manages infrastructure-based LTE-U and Wi-Fi networks (Fig.~\ref{fig:scenario}).
Every AP is required to first register with the spectrum manager and obtain permission to operate. 
Moreover, they need to follow any instructions (like power/channel assignments) from the spectrum manager. 
Therefore, it is natural to propose fairness monitoring as part of the spectrum manager's extended functionality and responsibility. 

The fairness monitoring procedure is as follows.
First, the spectrum manager assigns a reasonable duty cycle limit to a LTE-U AP based on the current channel usage, which would serve as input to the duty cycle adaptation algorithm. 
Then the spectrum manager collects information from a local Wi-Fi AP for duty cycle estimation without interrupting its normal network operations.
If it is decided that the LTE-U AP is misbehaving (i.e., violating the rule in Eq.~(\ref{eq:duty_cycle})), it will be punished accordingly (e.g., temporary suspension)\footnote{Designing an appropriate punishment scheme is out of the scope of this paper and left as future work.}. 

We assume that the selected Wi-Fi AP can be configured to always physically senses the medium (regardless of  RTS/CTS messages sent by other nodes), and try to receive every Wi-Fi packet during the monitoring process.
We also assume that it reports requested information honestly; robust duty cycle estimation against possible misreporting is left as future work.
The start time and period of LTE-U cycles are honestly reported by LTE-U APs to the spectrum manager, since they have no incentives to misreport. 
But the actual duty cycle in each cycle is not reported due to signaling overhead, or can be easily misreported to avoid punishment.

\section{Duty Cycle Estimation and Misbehavir Detection}\label{sec:monitoring}
In this section, we discuss the information collected by the selected Wi-Fi AP, and present our duty cycle estimation and misbehavior detection schemes.

\subsection{PHY Layer Observations}
We first discuss what a Wi-Fi AP (as the observer node) that is interfered by the target LTE-U AP would observe at the PHY layer.
As shown in Fig.~\ref{fig:CSMA_CA_access_cycle}, normal Wi-Fi operations are characterized by frequent idle periods (e.g., SIFS, DIFS and BO periods), which can only be seen during LTE-U OFF time.
On the other hand, ON time will cause the observer node to detect busy medium for duration longer than any normal Wi-Fi packets.
By physically sensing the medium, the observer node can easily observe idle/busy periods, which will be useful for duty cycle estimation.

\begin{figure}[ht!]
	\begin{center}
    \includegraphics[width = 1\columnwidth]{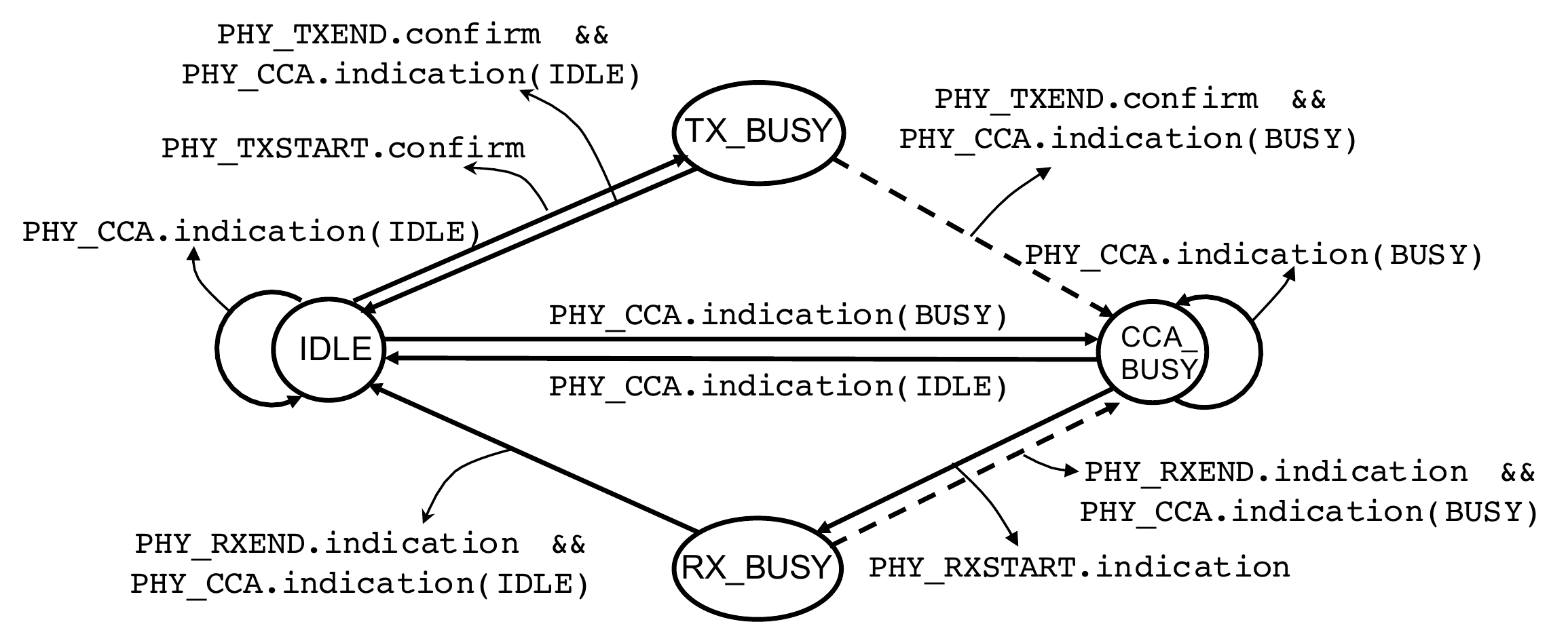}
    \caption{Wi-Fi PHY layer state machine.
    Transitions in dashed arrows are mainly caused by LTE-U transmissions.
    }
    \label{fig:phy_state_machine}
    	\end{center}
    	\vspace{-0.3cm}
\end{figure}

Now let us take a closer look at the PHY layer state machine of the observer node.
For duty cycle estimation, we are mainly interested in four PHY states: IDLE, CCA\_BUSY, TX\_BUSY and RX\_BUSY, as shown in Fig.~\ref{fig:phy_state_machine}. 
Transitions between the four states are triggered by medium busy/idle events as well as Tx/Rx events, and indicated by primitives that are already available at the MAC layer \rev{as per the Wi-Fi standard}.

Consider an observer node in IDLE.
When it starts transmission, the {\tt PHY\_TXSTART.confirm} primitive is issued, and it enters TX\_BUSY;
the end of transmission is indicated by {\tt PHY\_TXEND.confirm}. 
If other node transmits, the medium will become busy, and the {\tt PHY\_CCA.indication(BUSY)} primitive will be issued. 
The observer node will go to CCA\_BUSY, looking for valid Wi-Fi preambles. 
If a valid Wi-Fi preamble and header are received, 
 {\tt PHY\_RXSTART.indication} is issued, and the observer node goes to  RX\_BUSY.
It stays there till the end of predicted duration (indicated by {\tt PHY\_RXEND.indication}). 
If the preamble or header is missed, it stays in CCA\_BUSY for a period equal to the Wi-Fi packet transmission duration.
Under normal conditions, the observer node is expected to return to IDLE  after transmission or reception of a Wi-Fi packet.
Note that it is possible that two Wi-Fi packets collide, and the node (i.e., the observer node) transmitting a shorter packet detects busy medium (CCA\_BUSY) immediately after transmission (TX\_BUSY).
But this rare case can be ignored safely.

Since a LTE-U AP may transmit anytime without notifying Wi-Fi nodes, the PHY layer state machine is impacted in the following way.
If ON time starts when the medium is idle, the observer node will immediately transit from IDLE to CCA\_BUSY, and stay there till the end of ON time, since no Wi-Fi preamble or header will be received.
If ON time starts during TX\_BUSY or RX\_BUSY, the observer node operates as usual, since it cannot immediately detect the presence of LTE signals.
But it enters CCA\_BUSY instead of IDLE after Tx/Rx is over, since ON time is longer than a Wi-Fi packet.  

As we can see, the observer node can indeed observe idle ($\mathcal{I}$) and busy ($\mathcal{B}$) periods that appear alternately very easily, as well as Tx/Rx duration, by keeping track of related primitives available at the MAC layer.
Although the time spent in the three busy states (i.e., CCA\_BUSY, TX\_BUSY and RX\_BUSY) is counted towards busy periods, a busy period will be labeled differently with $\mathcal{B}_{tx}$ or $\mathcal{B}_{rx}$, if TX\_BUSY or RX\_BUSY is visited.

In practice, when a Wi-Fi AP receives a monitoring request from the spectrum manager, it starts recording observed busy periods for the requested time window chronologically. 
Let the $i$-th observed busy period be $(t_i, label_i, d_i, d_i')$, where $t_i$ is the start time, $label_i \in \{\mathcal{B},  \mathcal{B}_{tx},  \mathcal{B}_{rx}\}$ is the label, $d_i$ is the duration of that busy period, and $d_i'$ is the time spent in TX\_BUSY or RX\_BUSY, if $label_i = \mathcal{B}_{tx}$ or $\mathcal{B}_{rx}$ (otherwise $0$).
Note that in each busy period, either TX\_BUSY or RX\_BUSY may  be visited at most once, since it is not possible for a Wi-Fi node to transit directly from TX\_BUSY to RX\_BUSY (or vice versa) without visiting IDLE.
Finally, this Wi-Fi AP reports observed busy periods to the spectrum manager for duty cycle estimation.

\subsection{Duty Cycle Estimation}
\begin{figure}[ht!]
    \begin{center}
    \includegraphics[width = .8\columnwidth]{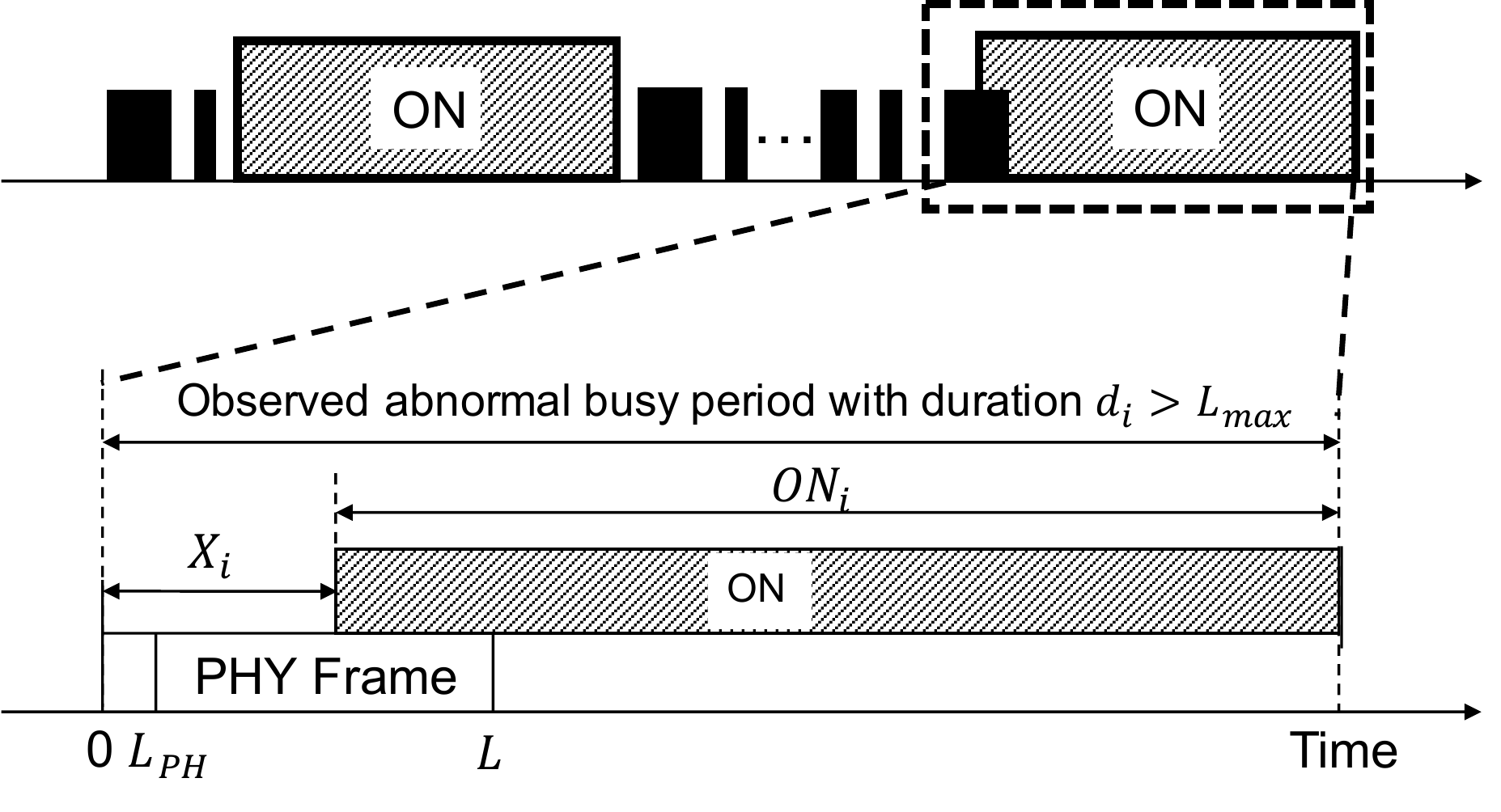}
    \caption{In addition to a continuous LTE-U ON period, an abnormal busy period may contain a portion of a Wi-Fi packet due to collision.}
    \vspace{-0.3cm}
    \label{fig:ON_overlap_with_packet}
    \end{center}
\end{figure}

Since the spectrum manager knows the start time and period of LTE-U cycles, it can focus on duty cycle estimation for each individual cycle of period $T$.
Although a busy period may be caused by either Wi-Fi or LTE-U transmissions, an \textit{abnormal} busy period with duration that is much longer than a Wi-Fi packet must contain a continuous LTE-U ON period.
In fact, it may also contain a portion of a Wi-Fi packet (Fig.~\ref{fig:ON_overlap_with_packet}), due to LTE-U/Wi-Fi collision.

Denote the set of abnormal busy periods in a cycle as $S'=\{(t_i, label_i, d_i,d'_i):label_i \in \{\mathcal{B}, \mathcal{B}_{tx}, \mathcal{B}_{rx} \}, d_i > L_{\max}\}$, where $|S'|=m$ and $L_{\max}$ is the maximum transmission duration of Wi-Fi packets. 
For convenience, elements in $S'$ are relabeled from $1$ to $m$.  
As mentioned earlier, LTE-U ON time may or may not overlap with an ongoing Wi-Fi transmission (Fig.~\ref{fig:ON_overlap_with_packet}).
If the LTE-U AP starts transmission before any Wi-Fi node transmits during an idle period, the observer node will sense the medium to be \rev{busy} and back off. 
In this case, the observed abnormal busy period contains only CCA\_BUSY duration\footnote{Note that it is possible that the ON period overlaps with the Wi-Fi preamble and header, or with two Wi-Fi packets that happen to collide, in which case RX\_BUSY is not visited, and the observer node stays in CCA\_BUSY.
In the first case, we still have $ON_i \approx d_i$, since the Wi-Fi preamble and header length (in 10s of $\mu$s) is much smaller than the ON period (in 10s of ms). 
We intentionally ignore the second case, since the probability is small (especially when RTS/CTS is enabled). 
}, which is equal to the ON duration.
In other words, we have $ON_i = d_i$, if $label_i = \mathcal{B}$.

However, if ON time starts during an ongoing Wi-Fi transmission, then the observed abnormal busy period includes a portion of the Wi-Fi packet at the beginning, followed by an ON period. 
There exist two cases here. 
In the first case, the observer node is currently transmitting a packet of duration $L$.
In the second case, the observer node has detected a valid Wi-Fi preamble and header (of length $L_{PH}$), and is currently receiving the payload of predicted duration $(L-L_{PH})$ (which can be inferred from the LENGTH field in the header). 
In both cases, it will stay in TX\_BUSY or RX\_BUSY for duration $L$ before going to CCA\_BUSY.

Denote the portion of the Wi-Fi packet that is not overlapping with the ON period as $X_i$. 
We can see that
\begin{equation*}
ON_i = d_i - X_i, \label{eq:observed_ON_time}
\end{equation*}
where $X_i \in [0, L]$ if $label_i = \mathcal{B}_{tx}$, and $X_i \in [L_{PH}, L]$ if $label_i = \mathcal{B}_{rx}$.
\rev{Due to lack of information about $X_i$}, we model it as a uniform random variable when $label_i$ is  $\mathcal{B}_{tx}$ or  $\mathcal{B}_{rx}$.
Then we have
\vspace{-0.2cm}

\footnotesize
\begin{equation*}
ON_i \sim 
\begin{cases}
d_i, & \text{ if } label_i = \mathcal{B}\\
Unif [d_i - L, d_i], & \text{ if } label_i = \mathcal{B}_{tx} \\
Unif [d_i - L, d_i - L_{PH} ], & \text{ if } label_i = \mathcal{B}_{rx},
\end{cases} \label{eq:ON_period_model}
\end{equation*}
\normalsize
where $L=d'$ in our model.
It is reasonable to assume that $\{X_i\}$ are independent of each other.
The spectrum manager adopts the following estimator for $ON_i$, 
\vspace{-0.2cm}

\footnotesize
\begin{equation*}
\hat{ON}_i =
\begin{cases}
d_i, & \text{ if } label_i = \mathcal{B}\\
d_i - \frac{1}{2} d'_i, & \text{ if } label_i = \mathcal{B}_{tx}\\
d_i - \frac{1}{2}(d'_i+L_{PH}), & \text{ if } label_i = \mathcal{B}_{rx}
\end{cases}, \label{eq:expected_ON}
\end{equation*}
\normalsize
and $\mathbb{E}[\hat{ON}_i] = ON_i$, which means that $\hat{ON}_i$ is an unbiased estimator. 
The spectrum manager estimates the duty cycle $\hat{\alpha}$ as follows,
\begin{align*}
\hat{\alpha} = \frac{1}{T}\sum_{i=1}^m \hat{ON}_i, \label{eq:estimated_alpha}
\end{align*}
which is also an unbiased estimator, since $\mathbb{E}[\hat{\alpha}] = \alpha$.

Note that although in practical LTE systems, one subframe of $1$ms duration is usually the minimum time unit for resource allocation, we do not make this assumption, and stay with the general case.  
Nevertheless, such knowledge can potentially increase estimation accuracy, and our logic is still applicable to this special case.

\subsection{Misbehavior Detection}\label{sec:misbehavior_detection}
After obtaining $\hat{\alpha}$, the spectrum manager needs to determine whether the LTE-U AP violates the rule in Eq.~(\ref{eq:duty_cycle}). 
%
The detection scheme is as follows,
\begin{equation*}
Result = 
\begin{cases}
\text{Violated}, & \text{ if } \hat{\alpha} > (1+\gamma) \alpha_{\max} \\
\text{Not violated}, & \text{ Otherwise },
\end{cases}
\end{equation*}
where $\gamma \geq 0$ is a parameter set by the spectrum manager.

Its performance is measured by \textit{probability of detection} $P_d$ and \textit{probability of false alarm} $P_{fa}$, i.e.,
\begin{align*}
P_d(\alpha, \gamma) &= \text{Pr}(\hat{\alpha} > (1+\gamma) \alpha_{\max} | \alpha > \alpha_{\max}) \\
P_{fa}(\alpha, \gamma) & = \text{Pr}(\hat{\alpha} > (1+\gamma) \alpha_{\max}| \alpha \leq \alpha_{\max}).
\end{align*}

To understand $P_d$ and $P_{fa}$, we consider the example in Fig.~\ref{fig:LTE_U_duty_cycle}, in which the LTE-U AP transmits continuously for $ON_{\max}$ of $20$ms and pauses for a short period of $2$ms, before transmitting again for another duration equal to (or less than) $ON_{\max}$.
We set $\alpha_{\max}$ to $0.5$.
We consider the worst case that each ON period overlaps with a Wi-Fi packet of length $L_{\max}$, and ignore $L_{PH}$ when $label_i=\mathcal{B}_{rx}$ for simplicity. 
Then we have
\vspace{-0.2cm}

\footnotesize
\begin{align*}
\hat{\alpha} &= \frac{1}{T} \sum_{i=1}^m \left(d_i - \frac{1}{2}L_{\max} \right)  = \frac{1}{T} \sum_{i=1}^m \left(ON_i + X_i - \frac{1}{2}L_{\max} \right)  \nonumber \\
& =  \alpha + \frac{1}{ T} \left(\sum_{i=1}^m X_i - \frac{1}{2}mL_{\max}  \right),
\end{align*}
\normalsize
where $X_i \in Unif[0, L_{\max}]$ and $m=\lceil\frac{\alpha T}{ON_{\max}} \rceil$. 

Define $X_i'=\frac{X_i}{L_{\max}}\in Unif[0,1]$. 
Then the sum of $m$ i.i.d. random variables $Y=\sum_{i=1}^m X_i'$ follows the Irwin-Hall (or uniform sum) distribution, that is,
\begin{equation}
F_Y(y) = \text{Pr}(Y\leq y) = \frac{1}{m!} \sum_{k=0}^{\lfloor y \rfloor} (-1)^k \binom{m}{k} (y-k)^{m}, \label{eq:uniform_sum_cdf}
\end{equation}
\normalsize
and it has a mean of $\frac{m}{2}$ and a variance of $\frac{m}{12}$.
When $m$ is large, the distribution of $Y$ can be well approximated by a Gaussian distribution $N(\frac{m}{2}, \frac{m}{12})$. 
But $m$ may be small in our case, and  we will use Eq.~(\ref{eq:uniform_sum_cdf}).
Then the probability of $\hat{\alpha}>(1+\gamma)\alpha_{\max}$ is given by
\begin{align}
\text{Pr} & \left\{ \hat{\alpha}>(1+\gamma)\alpha_{\max} \right\} \nonumber \\
&= \text{Pr}\{ \alpha+ \frac{L_{\max}}{T} \left( Y - \frac{m}{2} \right) > (1+\gamma) \alpha_{\max} \} \nonumber \\
&= 1 - F_Y \left( \frac{m}{2} + \frac{T}{L_{\max}} [(1+\gamma)\alpha_{\max} - \alpha] \right). \label{eq:prob_of_detection}
\end{align}

\begin{figure}[t!]
    \centering
    \includegraphics[width = 1\columnwidth]{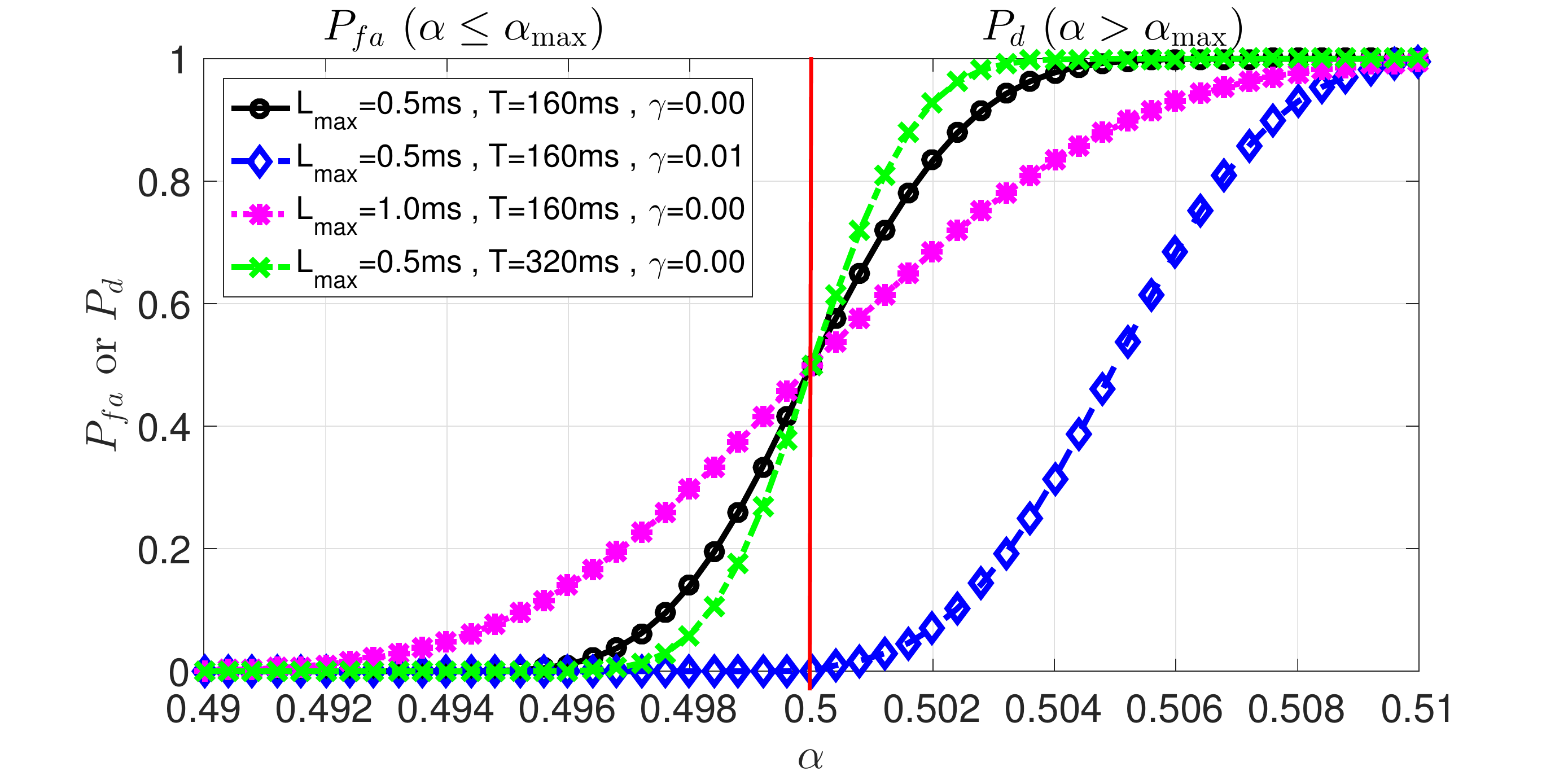}
    	\caption{$P_d$ and $P_{fa}$ as function of $\alpha$ with different values of $\gamma$, $L_{\max}$ and $T$, where $\alpha_{\max}$ is set to $0.5$.}
    	\label{fig:Pd_Fa_curve1_alpha}
\end{figure}

Note that if the true duty cycle $\alpha$ is greater than $\alpha_{\max}$, the probability in Eq.~(\ref{eq:prob_of_detection}) is $P_d$; otherwise, it becomes $P_{fa}$. 
For instance, with $L_{\max}=0.5$ms, $T=160$ms and $\gamma=0$ (i.e., the black curve in Fig.~\ref{fig:Pd_Fa_curve1_alpha}), if the LTE-U AP transmits with a duty cycle of $0.498$, the probability of mistakenly identifying that AP as misbehaving is $14.0\%$. 
If $\alpha=0.502$, the probability of correctly detecting that misbehaving user is $83.4\%$.

Next we study the impact of $\gamma$, $L_{\max}$ and $T$ on Eq.~(\ref{eq:prob_of_detection}).
In Fig.~\ref{fig:Pd_Fa_curve1_alpha}, the setting with $L_{\max}=0.5$ms, $T=160$ms and $\gamma=0$ is considered as the baseline. 
When $\gamma$ is increased to $0.01$, the curve is shifted to the right, which implies smaller $P_{fa}$ for any $\alpha \leq \alpha_{\max}$ but also smaller $P_d$ for any $\alpha > \alpha_{\max}$. 
Hence, it implies a tradeoff between $P_d$ and $P_{fa}$ when adjusting $\gamma$. 

Then we increase $L_{\max}$ to $1.0$ms while keeping other parameters the same with the baseline.
We can see an increase in $P_{fa}$ and a decrease in $P_{d}$, which means that larger Wi-Fi packets will adversely impact the detection performance of the proposed scheme. 
In contrast, when $T$ is increased to $320$ms, the overall detection performance is better, i.e., smaller $P_{fa}$ and larger $P_d$. 

\section{Evaluation}\label{sec:evaluation}
In this section, we evaluate the proposed duty cycle estimation and misbehavior detection schemes. 

\subsection{Simulation Setup}
We implement and evaluate the proposed schemes in \textit{ns3} \cite{ns3}, a widely used network simulator.
We consider the coc-channel coexistence of a LTE-U cell and a Wi-Fi network that consists of an AP and $20$ clients, all of which are located close to each other.
Simulation parameters are provided in Table~\ref{table:simulation_parameters}. 

Each Wi-Fi node has a full outgoing buffer of $1000$-byte UDP packets.
By adjusting the A-MSDU threshold \rev{(for all Wi-Fi nodes)}, variable maximum transmission duration of Wi-Fi (data) packets is obtained.
An adaptive but idealized, feed-back Wi-Fi rate control is used, where adjustments are made immediately upon feedback from the peer. 
PHY layer state information is obtained by tracing the PHY state machine of the Wi-Fi AP.

\begin{table}[htbp]
\caption{Simulation parameters}
\label{table:simulation_parameters}
\footnotesize
\centering 
\begin{tabular}{l c}
\hline
Parameter  	& 	Value \\ \hline
Wi-Fi standard 		& 802.11n (Mixed Format) \\
Channel 	& $20$MHz ($5170$-$5190$MHz) \\
Wi-Fi AP/client Tx power & $24$/$18$ dBm \\
CCA-CS/ED threshold & -$82$/-$62$dBm \\
Traffic model & Full buffer UDP \\
RTS/CTS	 & Disabled\\
Frame aggregation 	 & A-MSDU enabled\\
Min./max. continuous ON period & $6$/$20$ms \cite{lteUforum2015coexistence}\\
Idle gaps between ON periods & $2$ms \\
LTE-U cycle period ($T$) & $80$-$480$ms  \\
Max. Wi-Fi packet duration ($L_{\max}$) & $300$-$1100\mu$s \\
Max. duty cycle ($\alpha_{\max}$) & $0.5$\\
\hline
\end{tabular}
\end{table}

Since LTE-U performance (e.g., throughput) is not our concern in this work, we implement the LTE-U AP as a non-communicating device that switches on or off. 
We consider the typical duty cycling pattern in Fig.~\ref{fig:LTE_U_duty_cycle}, in which the LTE-U AP transmits continuously for $20$ms and pauses for $2$ms before transmitting for another $20$ms (or less) \cite{qualcomm2015}. 
Although the transmission duration a Wi-Fi packet can be up to $3$ms, it makes little sense for Wi-Fi data packets to be much larger than $1$ms when coexisting with LTE-U, since the idle gap is only $2$ms.
Consistent with  \cite{lteUforum2015coexistence}, we set $\alpha_{\max}$ to $0.5$.

\subsection{Duty Cycle Estimation}
\begin{figure}[!t]
\centering
\subfloat[]{\includegraphics[width=.48\columnwidth]{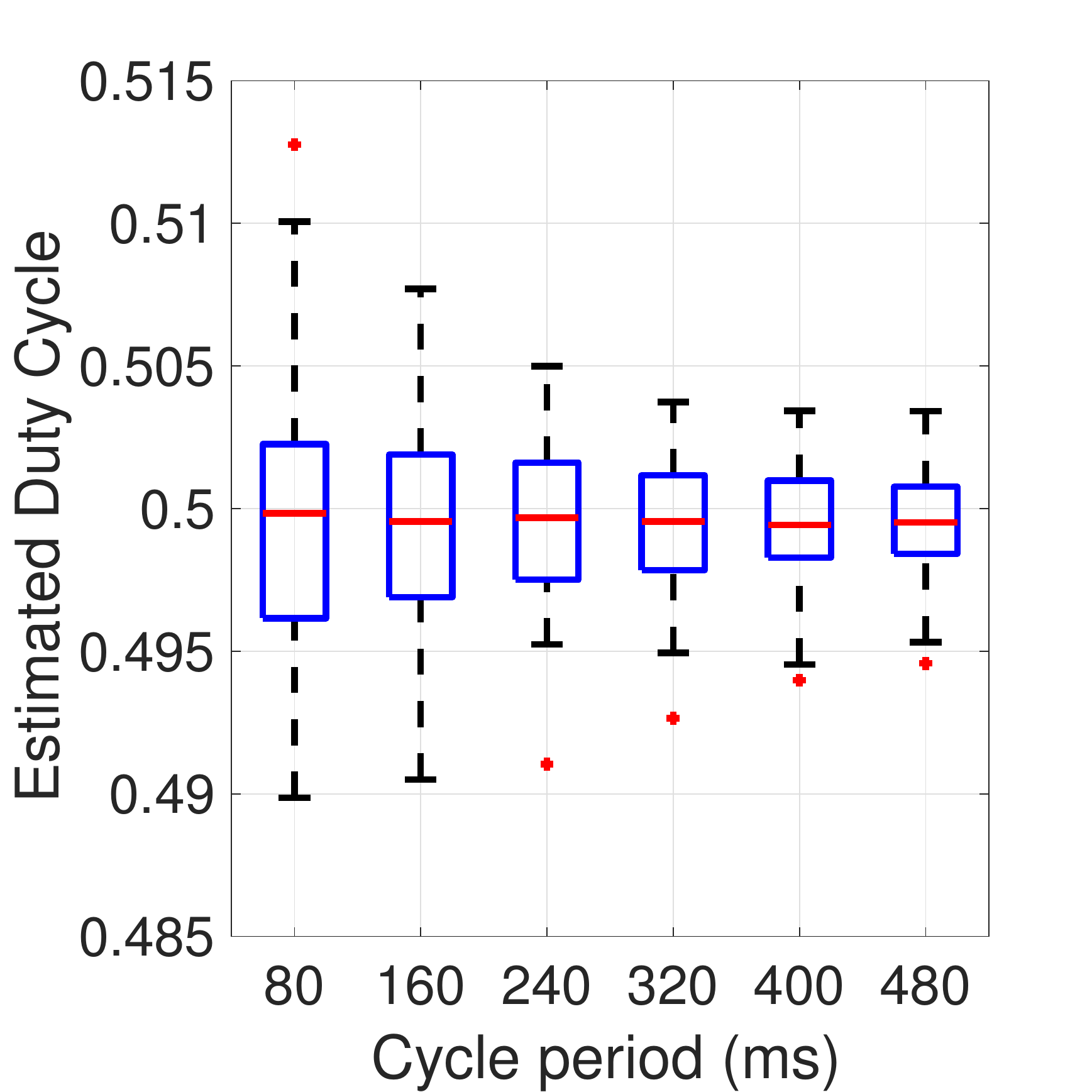}
\label{fig:impact_of_cyclePeriod}}
\hfil
\subfloat[]{\includegraphics[width=.48\columnwidth]{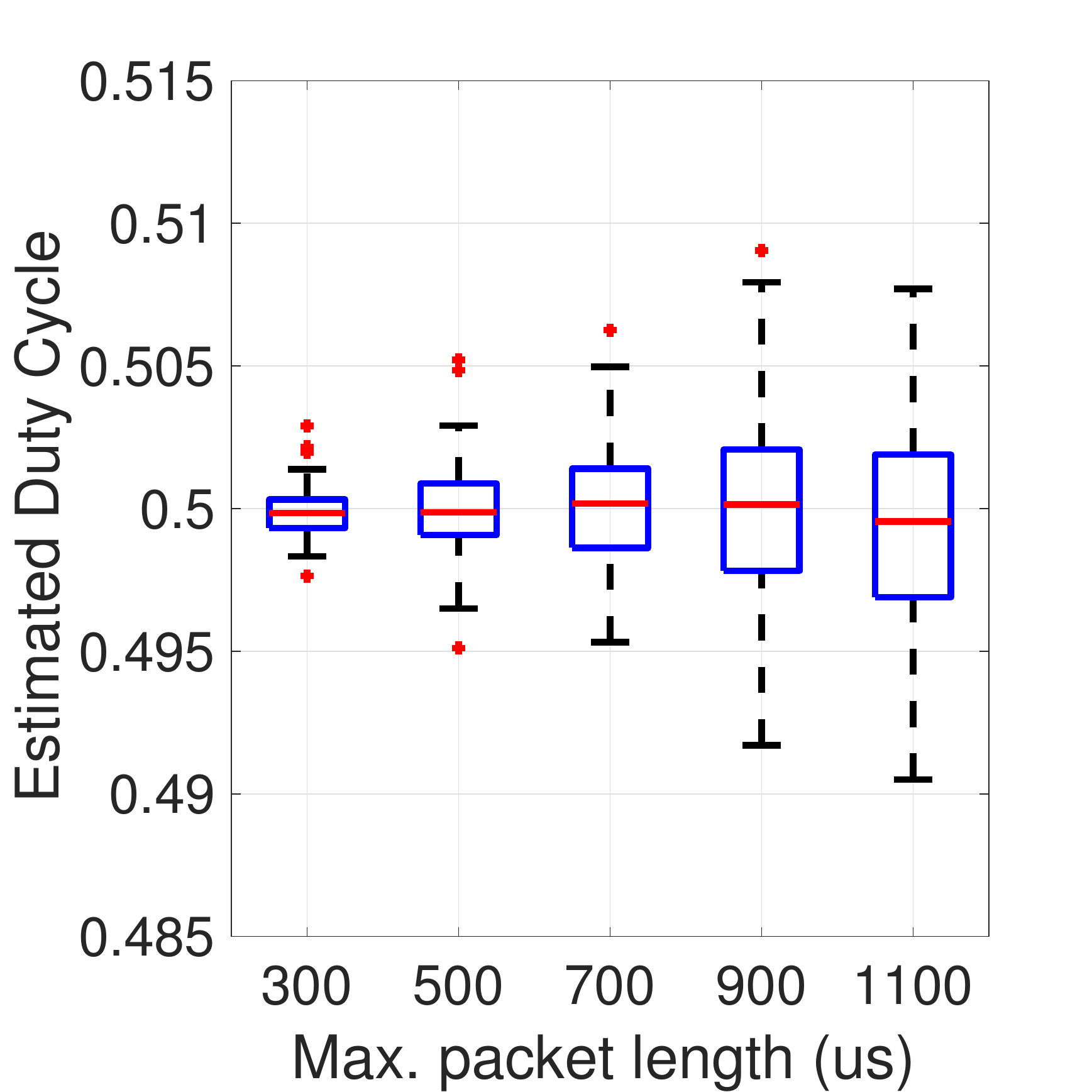}
\label{fig:impact_of_maxLength}}
\caption{Box plot of estimated duty cycle. (a) Impact of $T$ ($L_{\max}=1100\mu$s). (b) Impact of $L_{\max}$ ($T=160\mu$s). }
\label{fig:duty_cycle_estimation}
\vspace{-0.3cm}
\end{figure}

In this experiment, we evaluate the proposed duty cycle estimation scheme.
We first set $\alpha=0.5$, $L_{\max}\approx 1100\mu$s, and vary $T$ from $80$ms to $480$ms. 
Each experiment is repeated $100$ times for each setting, and results are shown in Fig.~\ref{fig:impact_of_cyclePeriod}. 
As we can see, the median of $\hat{\alpha}$ is very close to $\alpha$ for different $T$ values.
As $T$ increases, the deviation of $\hat{\alpha}$ is smaller, and the estimation is more accurate.
In all cases, $\hat{\alpha}$ is within $\pm 1\%$ of $\alpha$.


Then we fix $T$ to $160\mu$s, and vary $L_{\max}$ from $300\mu$s to $1100\mu$s.
Results from $100$ experiments are shown in Fig.~\ref{fig:impact_of_maxLength}.
We observe that $\hat{\alpha}$ is less  accurate as $L_{\max}$ increases, but $\hat{\alpha}$ is still within $\pm 1\%$ of the true duty cycle.
In practice, smaller Wi-Fi packets would help estimate $\alpha$ more accurately, but could  potentially decrease Wi-Fi throughput due to PHY and MAC overhead.

\subsection{Detection Performance}
\begin{figure}[ht!]
    \centering
    \includegraphics[width =1\columnwidth]{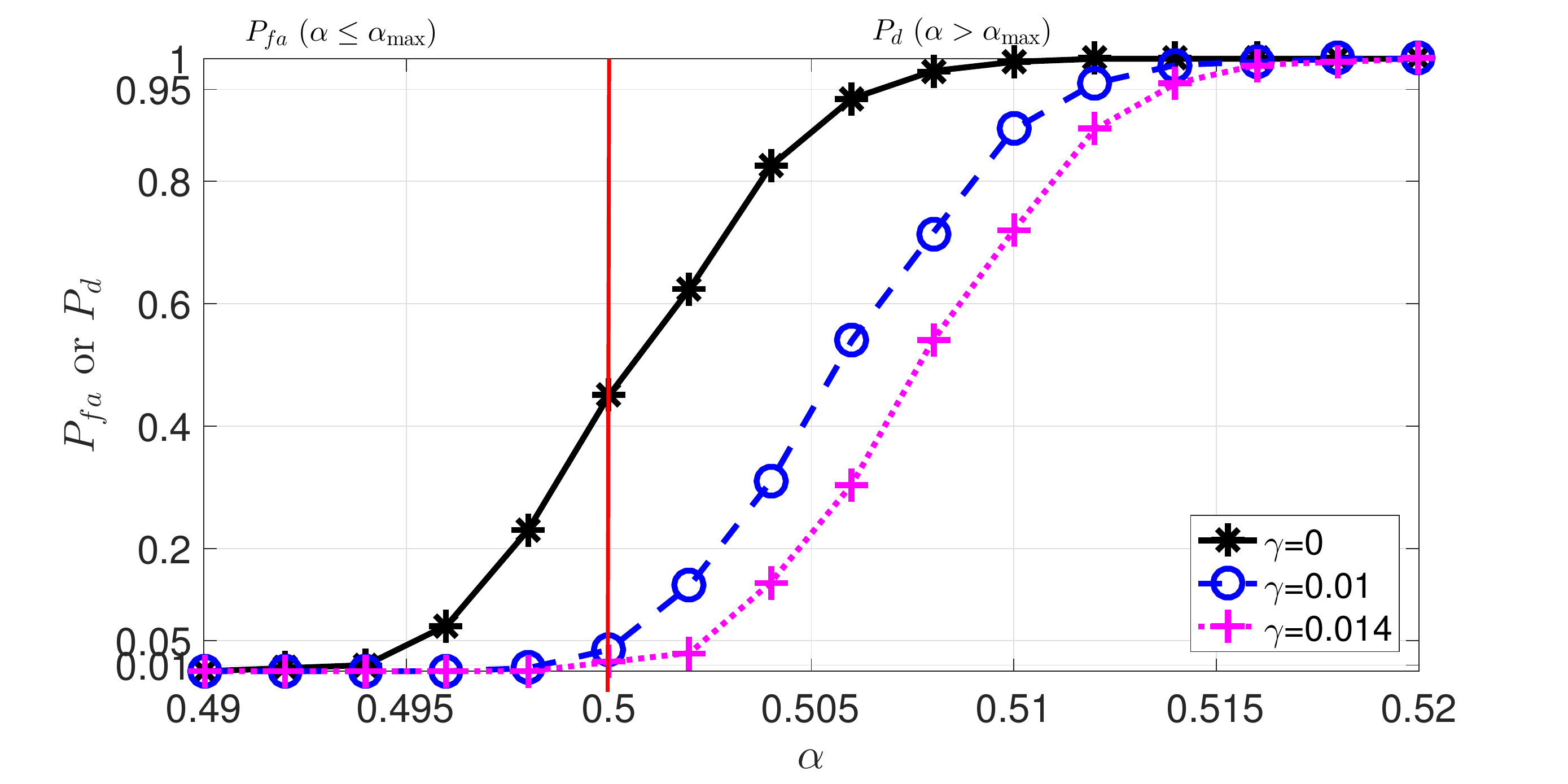}
    \caption{Performance of detecting duty cycling misbehavior with different $\gamma$ values, where $\alpha_{\max}=0.5$.
    Note that $\gamma=0$ is the baseline (in black curve), and the other two $\gamma$ values are chosen such that $P_{fa}$ is less than $5\%$ (in dashed blue curve) and $1\%$ (in dashed pink curve). }\label{fig:result_Pd_Fa_alpha}
    \vspace{-0.2cm}
\end{figure}

In this experiment, we evaluate the proposed misbehavior detection scheme in terms of $P_d$ and $P_{fa}$ with different choices of $\gamma$ values. 
We consider a typical cycle period of $160$ms \cite{qualcomm2015}, and set $\alpha_{\max}$ to $0.5$, $L_{\max}$ to $1100$ms. 
The true $\alpha$ is varied from $0.49$ and $0.52$. 
For each $\alpha$ value, the experiment is repeated $200$ times.
Results are shown in Fig.~\ref{fig:result_Pd_Fa_alpha}.

As we can see, with $\gamma=0$, $P_{fa}$ is as high as $45\%$ when $\alpha=0.5$, which is undesirable in practice. 
By setting $\gamma$ to $0.01$ or $0.014$, the spectrum manager can keep $P_{fa}$ under $5\%$ or $1\%$, but it also leads to a smaller $P_d$ for each $\alpha > \alpha_{\max}$. 
If the LTE-U AP transmits with a duty cycle higher than $0.514$, i.e., $2.8\%$ deviation from $\alpha_{\max}$, the proposed detection scheme with $\gamma=0.014$ can detect such misbehavior with a probability higher than $95\%$ while keeping $P_{fa}$ less than or equal to $1\%$.

\section{Conclusion}\label{sec:conclusion}
In this paper, we proposed a scheme that allows the spectrum manager to estimate LTE-U duty cycle via a local Wi-Fi AP.
We further proposed a scheme to detect possible misbehavior, and analyzed its performance in terms of detection and false alarm probabilities.
We implemented and evaluated the proposed schemes in \textit{ns3}. 
Our results have shown that the proposed schemes are able to estimate LTE-U duty cycle accurately, and detect misbehavior with a high detection probability and a low false alarm probability.


\bibliographystyle{IEEEtran}
\bibliography{references.bib}

\end{document}